\documentclass[fleqn,twoside]{article}
\usepackage{espcrc2}

\usepackage{graphics}

\usepackage{graphics,color,array,dcolumn}
\usepackage{calc}

\newcommand{\be}{\begin{equation}}
\newcommand{\ee}{\end{equation}}

\newcommand{\K}{{\kappa}}
\newcommand{\x}[1]{\xi_{#1}}
\providecommand{\la}{}
\renewcommand{\la}[1]{\lambda_{#1}}
\usepackage{amsmath}
\usepackage{amssymb}
\usepackage{xspace}

\newlength{\figurewidth}
\title{Gravity and Matter with Asymptotic Safety}

\author{Daniele~Perini\address[SISSA]{SISSA, via Beirut 4, I-34014
    Trieste, Italy
        and INFN, Sezione di Trieste, Italy}%
      \thanks{Electronic addess: \texttt{perini@he.sissa.it}}}
\begin{document}

\begin{abstract}
Building a consistent Quantum Theory of Gravity is one of the most
challenging aspects of modern theoretical physics. In the past
couple of years, new attempts have been made along the path of
``asymptotic safety'' through the use of Exact Renormalisation
Group Equations, which hinge on the existence of a non-trivial
fixed point of the flow equations. We will first summarize the
major results that have been obtained along these lines, then we
will consider the effect of introducing matter fields into the
theory. Our analyses show that in order to preserve the existence
of the fixed point one must satisfy some constraints on the matter
content of the theory. \vspace{1pc}
\end{abstract}

\maketitle

\section{Asymptotic Safety}
Quantisation of gravity has been one of the most intriguing and
fruitful fields of research in theoretical physics in the last
decades. The standard perturbation theory applied to the
Einstein-Hilbert Lagrangian,
\begin{equation}
S_{E-H}=\K\int\,\mathrm{d}^4x\sqrt{g}\left(2\Lambda-
  R\right),\label{eq:E-H}
\end{equation}
does not lead to a predictive theory because it requires an infinite
set of counterterms to cancel the divergences and therefore infinitely
many parameters should be determined experimentally. This is basically
due to the fact that it contains a negative-dimension coupling
constant. Such failure has brought to the quest for alternative ways
to standard field theory to quantise the metric.

In the last couple of years, though, a new line of investigation has
appeared in the literature which relies on nonperturbative methods. It
is based on the application of the Renormalisation Group (RG) to a
general coordinate invariant theory, through the use of Exact RG
Equations (ERGEs). If one finds that there exists a fixed point (FP)
of the RG flow which is UV attractive in a \emph{finite} number of
directions, then the theory is said to be \emph{asymptotically safe}
\cite{Weinberg:1979}, and is nonperturbatively renormalisable.
Consider the set of all quantum actions with running coupling
constants that possess a certain symmetry (in the gravity case it is
general covariance). A point in such a space is parameterized by the
infinite number of coupling constants. Suppose the theory allows for
an FP: the subspace of actions that flow towards it in the UV regime
makes up the UV critical surface. If this surface happens to be
finite-dimensional, then all actions lying on it will be located by a
finite number of couplings, and the UV limit may be taken in a
controlled way since the couplings will not blow up while approaching
it, therefore avoiding the divergences that are typical of
nonrenormalisable theories. In this way, the theory is predictive and
makes sense at all energy scales, thus it can be considered as a
fundamental one.

Asymptotic safety at the Gaussian FP (GFP), \emph{i.~e.}\ that special
point where all couplings vanish in the UV, is equivalent to
standard renormalisability along with asymptotic freedom, so this
feature is a generalisation of the usual renormalisability concept.

To see whether this scenario holds for some theory, one has
to write down the RG equations for the coupling constants, find a
UV attractive FP, if any are there, and finally calculate the
dimension of the critical surface.

\section{ERGEs and Gravity Theory}
Since we already know that the GFP does not serve the purpose of
asymptotic safety, the theory being perturbatively nonrenormalisable,
we cannot resort to perturbative techniques; rather, we have employed
ERGEs
\mbox{\cite{Polchinski:1984gv,Bagnuls:2000ae,Berges:2000ew,Wetterich:1993yh}},
which contain genuine nonperturbative information in spite of the
approximations that one necessarily has to make.

\begin{figure}[htb]
  
  \centering{\resizebox{\figurewidth}{!}
    {\includegraphics{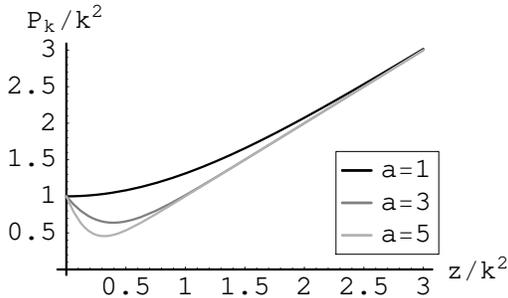}}\vspace*{-8mm}}
  \caption{\label{fig:Pk} Modified propagator $P_k(z)=z+R_k(z)$;  
    $a$ is a free parameter which determines the shape of $R_k$ (see
    Eq.~\eqref{eq:Rk}).}
\end{figure}

In Wetterich's formulation \cite{Wetterich:1993yh}, one considers
a scale-dependent action $\Gamma_k$, which describes the physics
at a typical energy scale $k$. It is a coarse-grained quantum
effective action, in Wilson's sense, which interpolates between
the classical action $S$ for $k\to\infty$ and the standard
effective action (the generator of the 1PI diagrams) for $k\to0$.
For the case of a single scalar field, the ERGE takes the form
\begin{equation}
  \label{eq:Wett}
  k\frac{\partial}{\partial k}\Gamma_k={1\over 2} {\rm Tr}\left[
  \left({\delta^2 \Gamma_k\over\delta\phi\delta\phi}+R_k\right)^{-1}
  k\frac{\partial}{\partial k}R_k\right].
\end{equation}
The trace is over momenta and $R_k$ is an IR cutoff entering the
classical action through a quadratic term, \mbox{$\Delta_k
S=\int\!\frac12\phi R_k \phi$}. Therefore this term modifies the
propagator $P_k$ of the low-momentum modes (w.r.t.\ the scale
$k$), as shown in \mbox{Fig.~\ref{fig:Pk}.}%
Effectively, these modes are suppressed in the functional integration,
as required in Wilson's formulation of the RG. The shape of $R_k$ is
controlled by an arbitrary parameter $a$, which will be needed later
on as a check of the consistency of the approximations. Explicitly, in
momentum space:
\begin{equation}
  \label{eq:Rk}
  R_k(z)=\frac{2a\,z\,e^{-2az/k^2}}{1-e^{-2az/k^2}}\quad
  \left(z=-\nabla^2\right).
\end{equation}
Numerical values will always be given for $a=\frac12$%
\footnote{In the exact theory there is no dependence on $a$, but the
  approximate results depend slightly on $a$ (see Sec.~5).}.

Eq. \eqref{eq:Wett} can be generalized to gravity
\cite{Reuter:1998cp}, adding gauge-fixing and ghost terms. Its
r.h.s.\ becomes a sum over second derivatives w.r.t.\ all fields.
Now the trace is over momenta and all quantum numbers, and $R_k$
is a matrix in the space of fields. Traces have been calculated
using standard heat-kernel techniques in the regime $k^2\gg R$.

Eq.~\eqref{eq:Wett} boils down to an infinite number of ordinary
first-order differential equations for the running couplings. To solve
this system a convenient way is to adopt a truncation, namely one
makes an ansatz such that $\Gamma_k$ is only made up of a suitable
subset of all the admissible operators. One then rescales the
dimensionful couplings with appropriate powers of $k$, ending up with
the set of dimensionless couplings $\{g_i\}$. An FP is solution of the
system \mbox{$\beta_i=\partial_t g_i=0$}, with $t=\log k$. To study
its attractivity, one can use the linearized
form of the flow equations,%
\begin{equation}
  \label{eq:lin.eq}
  \frac{\mathrm{d}g_i}{\mathrm{d}t}=M_{ij}\cdot g_j
  +\mathcal{O}\left(g^2\right),\quad
  M_{ij}=\left.\frac{\partial\beta_i}{\partial g_j}\right|_{FP},
\end{equation}
since $\beta_i=0$ at the FP. We shall call $M$ the stability matrix.
The solutions will be exponentials with the eigenvalues $\alpha_i$ of
$M$ at the exponent, \mbox{$\tilde g_i(t)\sim e^{\alpha_i t}$} (here
$\tilde g_i$ is an eigenvector of $M$, a linear combination of the
$g_i$'s). So the attractive directions for $k\to\infty$
will be those corresponding to eigenvalues of $M$ with a negative real
part\footnote{If there are vanishing eigenvalues, one must go beyond
  the linear order, but this will not be our case.}.

In the Einstein-Hilbert truncation, assuming \mbox{$\Gamma_k=S_{E-H}$}
with running $\Lambda$ and $\kappa$, one finds that there is a
non-Gaussian FP (NGFP) which is attractive in both directions
\cite{Lauscher:2001ya}. Introducing the dimensionless couplings
\mbox{$\hat{\Lambda}=k^{-2}\Lambda$} and
\mbox{$\hat{\kappa}=k^{-2}\kappa$}, one finds that their values at the
NGFP are
\mbox{$\hat{g}_*\equiv\left(16\pi\hat\kappa_*\right)^{-1}\!\!\approx
  0.344$} and \mbox{$\hat\Lambda_*\approx 0.339$}. Here and in the
following, the star denotes the values of the couplings at the FP.
Therefore, at least in this approximation, the theory is
asymptotically safe. The key point is to verify that the truncation
does not bring about fake results. Many checks have been made;
\emph{e.~g.}, in an exact treatment one would expect independence of
physical results from the cutoff function $R_k$, so the approximate
solution should at most show a mild dependence, and this is indeed
what happens \cite{Lauscher:2001cq,Lauscher:2001rz}. The major
achievement, however, is that the addition of an $R^2$-term, thus
considering a three-parameter truncation, does not change the results
significantly, so the results obtained in the two-parameter
Einstein-Hilbert truncation are trustworthy, and have a physical
meaning. This was not \emph{a priori} obvious: for instance, the GFP
disappears, so this was really an artifact of the truncation.

\section{Matter Fields}
\begin{figure}[htb]
  \centering{\resizebox{\figurewidth}{!}
    {\includegraphics{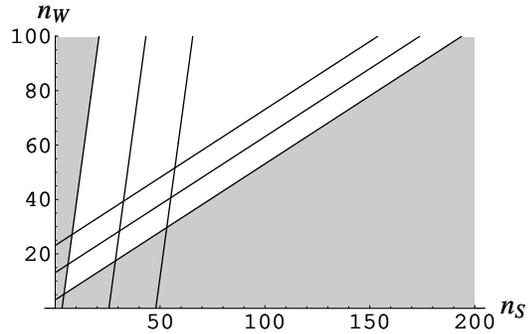}}
}
\caption{\label{fig:attr} NGFP existence regions for $n_{RS}=0$.
  The gray area shows the existence region of the NGFP for $n_M=0$,
  delimited by the two most external lines, whereas the white zone is
  the non-existence region. Going away from the gray area, the
  straight lines represent the boundaries of the regions for $n_M=10$
  and $n_M=20$, respectively. The existence region lies to the left
  and below these lines, being a translation of the gray area.}
\end{figure}

Now we extend the results found in \cite{Lauscher:2001ya} by including
matter fields. To begin with, we have considered
\mbox{$\Gamma_k=S_{E-H}$} extended with $n_S$ real scalar, $n_W$ Weyl,
$n_M$ Maxwell, and $n_{RS}$ (Majorana) Rarita-Schwinger fields, all
massless and minimally coupled \cite{Percacci:2002ie}.  Then we have
performed an analysis of the existence and attractivity of the NGFP
varying the number of matter fields. Some results are shown in
\mbox{Fig.~\ref{fig:attr}}%
for $n_{RS}=0$. The existence regions are bounded by the lines of
equations
\begin{subequations}  \label{eq:boundaries}
  \begin{gather}
    n_S+2n_M -2n_W-4n_{RS}+6.16=0, \label{eq:boundaries1} \\
    6.6 n_S -14 n_M -1.6 n_W+32n_{RS}-21=0. \label{eq:boundaries2}
  \end{gather}
\end{subequations}
Eq.~\eqref{eq:boundaries1} also discriminates between positive and
negative cosmological constant. $\hat\Lambda_*$ is positive when the
l.h.s.\ of Eq.~\eqref{eq:boundaries1} is.  Notice that it contains the
difference between the total numbers of bosonic and fermionic degrees
of freedom. In the existence region, the NGFP is always attractive in
both directions.

Now one can apply the bounds given by Eqs.~\eqref{eq:boundaries} to
see whether the coexistence of gravity with a certain matter theory is
still compatible with asymptotic safety. We have seen that popular GUT
$SU(5)$ and $SO(10)$ theories indeed are, yielding a positive or
negative $\hat\Lambda_*$ according to the symmetry-breaking pattern.
The bounds we found are difficult to evade, since a large number of
gauge bosons, as is the case for GUT theories, requires a large number
of fermions, so one should introduce many fermion families to violate
them.  As for supersymmetric contents of matter, they all lie below
the line of Eq.~\eqref{eq:boundaries1}, so in principle they would be
compatible with asymptotic safety; however, we found that in this
region numerical calculations are not reliable, depending quite
strongly on the cutoff function, so in the present context we cannot
say much about SUSY theories.

\section{Gravity with a Scalar Field}
{\centering
\begin{table*}[!t]\centering
  \caption{\label{table:stabmat}Stability matrix at the GMFP. The
    order of the variables is $\la0,\x0,\la2,\x2,\la4,\x4,\dots$}
  \begin{displaymath}
    \left(
      \begin{array}{ccccccccc}
        3.77 & -7.64 & -0.010  & -0.013 & 0 & 0 & 0 & 0 & \dots\\
        6.99 & -7.94 & 0.0031  & -0.019 & 0 & 0 & 0 & 0 & \dots\\
        0 & 0 & 5.77 & -7.64 & -0.063 & -0.078 & 0 & 0 & \dots\\
        0 & 0 & 6.99 & -5.94 & 0.019  & -0.11  & 0 & 0 & \dots\\
        0 & 0 & 0 & 0 & 7.77 & -7.64 &  -0.16  & -0.20 & \dots\\
        0 & 0 & 0 & 0 & 6.99 & -3.94 &  0.046  & -0.28 & \dots\\
        0 & 0 & 0 & 0 & 0 & 0 & 9.77 & -7.64 & \dots\\
        0 & 0 & 0 & 0 & 0 & 0 & 6.99 & -1.94 & \dots\\
        \dots & \dots & \dots & \dots & \dots & \dots & \dots & \dots & \dots
      \end{array}
    \right)
  \end{displaymath}
\end{table*}
}
An extension of the truncation considered in
\cite{Percacci:2002ie} would require a more detailed analysis of
the matter couplings. It seems impossible to conceive calculations
involving all admissible couplings present in a realistic matter
theory, so as a first step we have considered the simplest
example, that of a self-interacting scalar field
\cite{Percacci:2003jz}.  Aside from its role as a model for the
Higgs field in unified theories, a scalar field (the dilaton)
appears in many popular theories of gravity.  It can therefore
sometimes be regarded as part of the gravitational sector, rather
than the matter sector. This makes its properties especially
interesting in a gravitational context.

The class of running actions that we have considered is
\begin{multline}
\label{eq:class}
\Gamma_k[g,\phi]=\int\,\mathrm{d}^4x\sqrt{g}\;\cdot\\
\cdot\!\left(V(\phi^2)-
  F(\phi^2)R+{1\over2}g^{\mu\nu}\partial_\mu\phi\partial_\nu\phi\right),
\end{multline}
where the potential $V$ and the scalar-tensor coupling $F$
are arbitrary real analytic functions which can be expanded as a power
series in $\phi^2$:
\begin{subequations}\label{eq:power}
     \begin{align}
 V(\phi^2)&=\sum_{n=0}^\infty \tilde{\lambda}_{2n}
    \phi^{2n},\label{eq:powerV}\\
    F(\phi^2)&=\sum_{n=0}^\infty \tilde{\xi}_{2n}
    \phi^{2n}.\label{eq:powerF}
  \end{align}
\end{subequations}
As before, we introduce dimensionless couplings
{$\la{2n}=k^{2(n-2)}\tilde{\lambda}_{2n}$} and
\mbox{$\x{2n}=k^{2(n-1)}\tilde{\xi}_{2n}$}. We shall take into
account \emph{all} of these couplings.

The theory admits an NGFP where all couplings vanish, apart from
$\la0=2\hat\kappa\hat\Lambda$ and $\x0=\hat\kappa$. Since all matter
couplings vanish, we call it the ``Gaussian-Matter'' FP (GMFP). The
values of the cosmological and Newton's constant are only affected by
the presence of the scalar kinetic term, and turn out to be
\mbox{$\hat{g}_*\approx0.320$} and
\mbox{$\hat{\Lambda}_*\approx0.359$} at the GMFP. The
infinite-dimensional stability matrix is shown in
Table~\ref{table:stabmat}%
\footnote{Notice that in \cite{Percacci:2003jz} numerical values were
  given for $a=2$, so they are different from the ones we present
  here, but the parameter dependence is mild, see Sec.~5.}. %
One can see that it has an almost block-diagonal form, and the
diagonal blocks have a regular pattern.  The eigenvalues reflect this
regularity, being \mbox{$-2.08\pm4.38i$},
\mbox{$-0.08\pm4.38i$}, \mbox{$1.92\pm4.38i$},\dots%
\footnote{This is true if one considers $V$ and $F$ to be
polynomials. For a more detailed discussion see
\cite{Percacci:2003jz}.}
. The real parts of the eigenvalues increase by constant multiples
of two, so we can conclude that the critical surface has dimension
four, and the theory is again asymptotically safe at the GMFP.

Therefore, we can see that even though matter is ``Gaussian'', the
gravitational interactions produce significant changes to the
pure-scalar theory. For instance, the canonical dimension of the mass,
which is $\frac12 \tilde{\lambda}_2$, changes from 2 to $\sim\!0.12$
(after mixing with $\x2$) and the usual quartic coupling, $\la4$,
becomes now an irrelevant parameter, its eigenvalue having a positive
real part. The same pattern occurs for the other operators.

The analysis can be extended considering massless minimally coupled
fields of different spins added to this gravity-scalar system. The
outcome is that the NGFP is there, provided the matter content
satisfies the bounds of Eqs.~\eqref{eq:boundaries}, and it is
attractive in a possibly large number of directions. This could be a
solution to the well-known problem of the triviality of the scalar
theory; for a more detailed discussion of this issue
see~\cite{Percacci:2003jz}.

\section{Parameter Dependence}
As mentioned several times, a fundamental issue of this approach
is to test whether the approximations assumed in the truncation
are reliable. We can check whether physical results show a
dependence on the cutoff function by looking at their dependence
on the parameter a in Eq.~\eqref{eq:Rk}.

\begin{figure}[!t]
  \centering{\resizebox{\figurewidth}{!}
    {\includegraphics{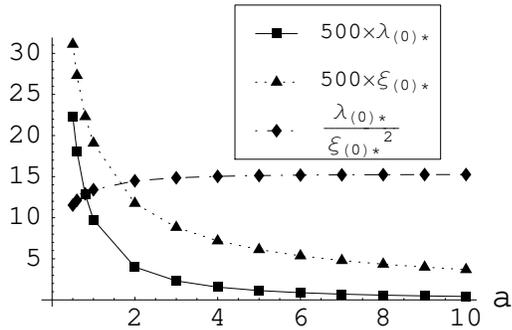}}
    }
\caption{\label{fig:on-shell}$\lambda_{0*}$, $ \xi_{0*}$, and
  $\lambda_{0*}/\xi_{0*}^2$ as functions of $a$. The first two
  functions are magnified to get them in the same range of values as
  the last one.}
\end{figure}

To give an instance, we can consider the ratio
$\lambda_{0*}/\xi_{0*}^2$, which is the inverse of the on-shell
action, up to numerical factors. It is therefore an observable
quantity, so in an exact treatment of our problem its value should
be independent of $a$. The state of the art is
depicted in Fig.~\eqref{fig:on-shell}.%
It is remarkable that while $\lambda_{0*}$ and $ \xi_{0*}$ display
quite a substantial dependence on $a$, the ratio
$\lambda_{0*}/\xi_{0*}^2$ is almost $a$-independent, giving an
encouraging hint towards the reliability of the truncation. The
stronger dependence of the latter quantity on $a$ for $a\to 0$ is
due to the fact that in this limit $R_k$ becomes a constant, so it
not trustworthy anymore as an IR cutoff.

Other cutoff-independent quantities are for instance the
eigenvalues of $M$, which show a reasonably mild dependence on
$a$ as well.

These results are in accordance with those of pure gravity
\cite{Lauscher:2001cq,Lauscher:2001rz}, from which they only differ
slightly because of the presence of the scalar field.

\section{Conclusions}
In this talk we have presented the concept of asymptotic safety and
reviewed the literature concerning its application to gravity theories
in a nonperturbative context, with the use of ERGE. We have seen that
the addition of massless, minimally coupled matter of different spins
to the theory, which in principle might spoil these beautiful
properties, can still yield an asymptotically safe theory, provided
one satisfies some weak bounds on the matter content.  The analysis of
gravity coupled to a single scalar field with an arbitrary potential
and coupling to the Ricci scalar, including infinitely many couplings,
shows that the theory is asymptotically safe at a ``Gaussian-Matter''
FP. In this case the scalar sector allows for a perturbative
treatment, whereas the gravity part is thoroughly non perturbative.
The canonical dimensions of the pure-scalar theory are significantly
changed by the gravitational corrections; this picture may ultimately
yield a solution of the triviality issue in the scalar theory.  The
reliability of the approximations has also been checked, giving
satisfying results. Therefore, this scenario has the potentiality of
giving a consistent field theoretical description of gravity and
matter.


\bibliographystyle{unsrt} \bibliography{database}

\end{document}